\documentclass{jfm}
\usepackage{color}
\usepackage{CJKutf8}
\usepackage{graphicx}
\usepackage{newtxtext}
\usepackage{newtxmath}
\usepackage{natbib}
\usepackage{hyperref}
\hypersetup{
    colorlinks = true,
    urlcolor   = black,
    citecolor  = blue,
}

\newcommand{\RomanNumeralCaps}[1]
\linenumbers
\title{Erodible bed turbulence modulation driven by transition between longitudinal and transverse bedforms at varying Shields numbers}

\author{Yinghaonan Lei\aff{1}, Ping Wang\aff{1}
 \and Xiaojing Zheng\aff{2}
 \corresp{\email{xjzheng@xidian.edu.cn}}}

\affiliation{
\aff{1}Center for Particle-laden Turbulence, Lanzhou University, Lanzhou 730000, PR China
\aff{2}Research Center for Applied Mechanics, Xidian University, Xi'an 710071, PR China
}

\begin{document}
\maketitle

\begin{abstract}
The mechanism of turbulence modulation in particle-laden flow over erodible beds remains an open question. Using particle-resolved direct numerical simulations, this study realises a longitudinal-to-transverse bedform transition by varying the Shields number, revealing non-monotonic modulation of near-wall turbulence. At low Shields numbers, streamwise sediment ridges generate form-induced streaks that produce a distinct secondary peak in the premultiplied energy spectra, exceeding the conventional near-wall turbulent peak and enhancing the turbulent kinetic energy. As the Shields number increases, saltation intensifies and disrupts these structures, causing the secondary peak to vanish in the streamwise direction and weaken in the spanwise direction, thereby suppressing turbulence. Proper orthogonal decomposition of the bed surface reveals a redistribution of modal contribution from a single dominant mode to higher-order modes, with longitudinal features persisting as remnants, directly linking bedform evolution to turbulence modulation.
\end{abstract}


{\bf MSC Codes }  {\it(Optional)} Please enter your MSC Codes here

\section{Introduction}
\label{sec:introduction}
\noindent \par Particle-laden turbulent flows over erodible beds are ubiquitous in aeolian \citep{zhengModulationTurbulenceSaltating2021} and fluvial \citep{garcia2008sedimentation} settings. Unlike smooth-wall flows, fluid shear over an erodible bed drives particle creep, saltation and suspension \citep{jiDirectNumericalSimulation2013}, directly modulating near-wall turbulence, and generates longitudinal \citep{zgheibDirectNumericalSimulation2018,schererRoleTurbulentLargescale2022} or transverse \citep{yalinMechanicsSedimentTransport1977a,bennettFluidSedimentDynamics1998} bedforms, further complicating the flow. Although turbulence modulation by particles over smooth walls exhibits multi-parameter dependence \citep{brandtParticleLadenTurbulenceProgress2022}, how near-wall mobile and bed particles jointly modulate turbulence over erodible beds remains an open question.

\par Most existing studies neglect or cannot resolve bedforms. \cite{zhengModulationTurbulenceSaltating2021} used wall-resolved LES with a hydraulically smooth wall and splash functions for particle--wall interactions, finding suppressed near-wall streamwise velocity fluctuations but enhanced vertical and spanwise fluctuations, attributed to splash-induced redistribution of turbulent kinetic energy from the streamwise to the vertical and spanwise components. \cite{zhangSimultaneousPIVPTV2008} and \cite{liBoundarylayerTurbulenceCharacteristics2012a} both reported turbulence enhancement in the saltation layer in wind-tunnel experiments, attributing it respectively to particle wakes, collision-induced vertical momentum injection, and increased effective roughness; bedforms could not be resolved in either study. \cite{liuExperimentalInvestigationEffects2022} decoupled the contributions of settling, collision and splash in a wind tunnel, finding that collision and splash predominantly suppress near-wall turbulence, though the sand bed was flattened after each run to exclude the effect of bedforms. Using particle-resolved direct numerical simulation (PR-DNS), \cite{jiDirectNumericalSimulation2013} found that, relative to a fixed rough wall, all three fluctuation-component peaks above an erodible bed were suppressed, while no pronounced bedforms were reported.

\par A few studies have examined the influence of bedform type on turbulence modulation, yet conclusions remain divided. \cite{raudkiviStudySedimentRipple1963} first measured turbulence over a rippled bed, finding the strongest fluctuations in the separated shear layer downstream of the lee side in water-tunnel experiments. \cite{bennettFluidSedimentDynamics1998} found that low-relief bedforms on a mobile bed enhance turbulence intensities relative to a fixed bed through lee-side wake formation, with increased equivalent sand roughness. \cite{revil-baudardTurbulenceModificationsInduced2016} similarly observed that a mobile sediment bed increases the turbulence kinetic energy (TKE) across the entire boundary layer. Using PR-DNS, \cite{vowinckelFluidParticleInteraction2014} varied the number of mobile particles over the fixed particle bed and found that, relative to the rough wall, ridge-like and dune-like structures self-organized by these particles enhance all Reynolds stress components by introducing spatial inhomogeneity. \cite{schererRoleTurbulentLargescale2022} adopted a smooth-wall baseline and found that near-wall particle transport over streamwise sediment ridges disrupts the self-sustaining cycle, suppressing the peak streamwise fluctuation. Thus, the diversity of bedform types and reference baselines precludes direct comparison of their contributions.
\par In summary, the diversity of particle scales, flow parameters, bedform types and baselines precludes direct comparison across existing studies. In erodible two-phase flows, \cite{shields1936application} introduced the Shields number $\theta = {\tau_w}/{(\rho_f R |\boldsymbol{g}| D_p)}$ to quantify the fluid shear driving strength on particles, where $\tau_w$ is the wall shear stress, $R = \rho_p/\rho_f - 1$ is the submerged specific gravity, $\rho_p$ and $\rho_f$ are the particle and fluid densities respectively, $\boldsymbol{g}$ is the gravitational acceleration and $D_p$ is the particle diameter. This parameter controls both particle motion and bedforms \citep{dey2014fluvial}, making it the key control parameter of erodible systems. Accordingly, this paper employs PR-DNS based on the immersed boundary method (IBM) to simulate incompressible particle-laden flow in an open channel over an erodible bed, reproducing different longitudinal and transverse bedforms by varying $\theta$ to reveal the mechanism of turbulence modulation under the combined action of mobile particles and bedforms. Section~2 introduces the numerical methodology. Section~3 presents the results. Concluding remarks are given in Section~4.
\section{Numerical Methodology and Simulation Framework}
\label{sec:Numerical_methodology_and_simulation_framework}
\noindent The governing equations for three-dimensional incompressible Newtonian fluid flow are:
\begin{equation}
\left\{
\begin{aligned}
\nabla \cdot \boldsymbol{u}_f &=0 \\
\frac{
\partial \boldsymbol{u}_f}{\partial t}
+\left(\boldsymbol{u}_f\cdot \nabla \right)\boldsymbol{u}_f
& =
-\frac{1}{\rho_f} \nabla p_e
-\frac{1}{\rho_f} \nabla p
+\nu_f \nabla^2 \boldsymbol{u}_f+\boldsymbol{f}_{pf}
\end{aligned}
\right.
\label{equ:ns}
\end{equation}
\noindent where $\boldsymbol{u}_f=[u_f,v_f,w_f]^T$ is the fluid velocity vector (superscript $T$ denotes transpose), $\boldsymbol{f}_{pf}$ is the particle feedback force, $p_e$ is the constant driving pressure, $p$ is the modified pressure \citep{breugemSecondorderAccurateImmersed2012}, and $\nu_f$ is the fluid kinematic viscosity. Periodic boundary conditions are applied in the streamwise $x$ and spanwise $z$ directions with a no-slip bottom wall and a free-slip top boundary. Spherical particles of diameter $D_p$ and density ratio $\rho_p/\rho_f=2.65$ are governed by:
\begin{equation}
\left\{
\begin{aligned}
\rho_p V_p \frac{\mathrm{d} \boldsymbol{u}_p}{\mathrm{d} t}
&=\boldsymbol{F}_{fp}
+\left(\rho_p-\rho_f\right) V_p \boldsymbol{g}
+\boldsymbol{F}_c\\
I_p \frac{\mathrm{d} \boldsymbol{\omega}_p}{\mathrm{d} t}
&=\boldsymbol{M}_{fp}
+\boldsymbol{M}_{c}
\end{aligned}
\right.
\label{equ:par_motion}
\end{equation}
\noindent where $\boldsymbol{u}_p=[u_p,v_p,w_p]^{T}$ and $\boldsymbol{\omega}_p=[\omega_{p,x},\omega_{p,y},\omega_{p,z}]^T$ are the translational and angular velocity vectors at the particle center, $V_p$ and $I_p$ are the particle volume and moment of inertia, $\boldsymbol{F}_{fp}$ and $\boldsymbol{M}_{fp}$ are the hydrodynamic force and torque, and $\boldsymbol{F}_c$ and $\boldsymbol{M}_c$ are the collision force and torque computed via the adaptive collision time model \citep{biegertCollisionModelGrainresolving2017}. The normal and tangential restitution coefficients and friction coefficient are $e_{n,d}=0.97$, $e_{t,d}=0.39$ and $\mu_c=0.15$, respectively. $\boldsymbol{f}_{pf}$, $\boldsymbol{F}_{fp}$ and $\boldsymbol{M}_{fp}$ are computed via IBM \citep{breugemSecondorderAccurateImmersed2012}. Details can be found in our previous work \citep{zhuParticleResolvedSimulation2022}.
\begin{figure}
    \centering
    \includegraphics[width=1\linewidth]{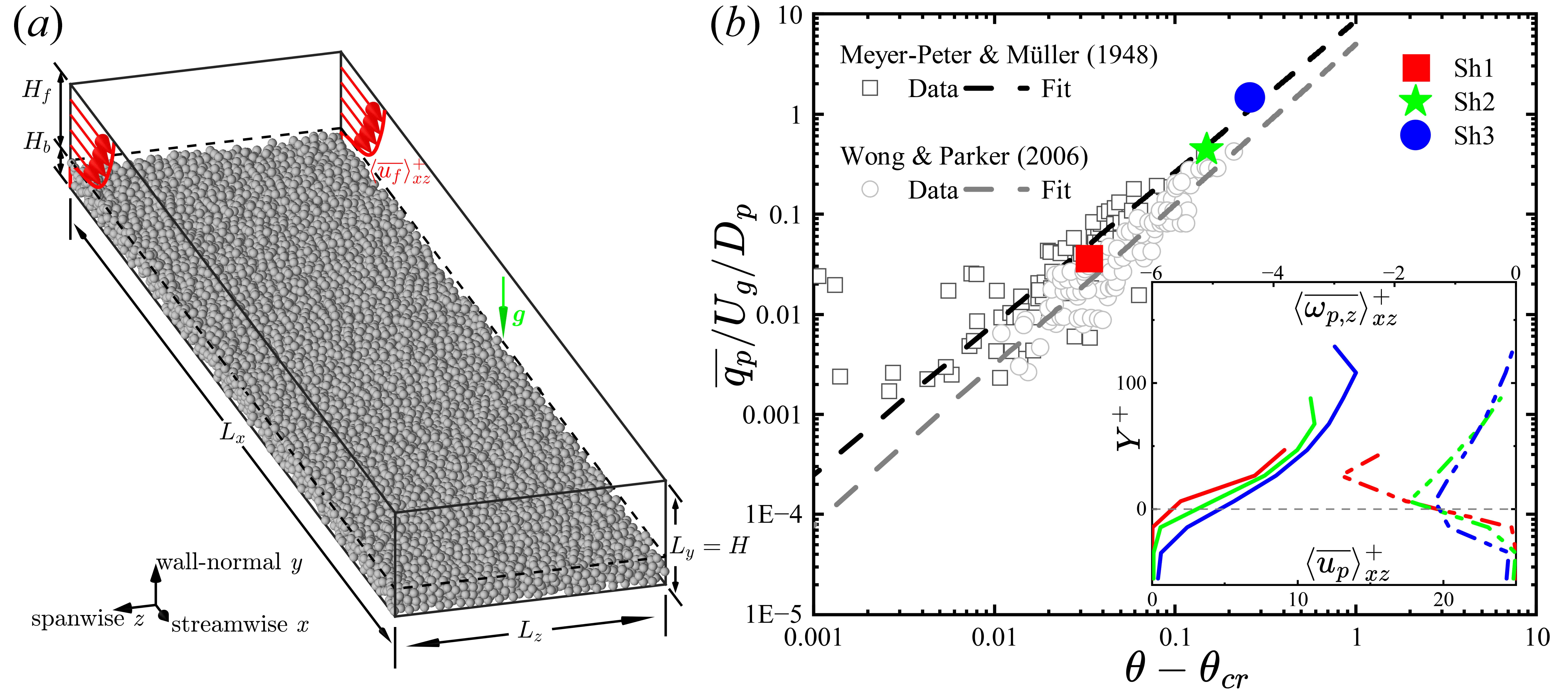}
    \caption{(a) Simulation setup for Sh0. (b) Normalized particle volume flux $\overline{q_p}/U_g/D_p$ versus $\theta$ for Sh1--Sh3. The inset shows $\langle \overline{u_p}\rangle_{xz}^+$ (solid, lower axis) and $\langle \overline{\omega_{p,z}}\rangle_{xz}^+$ (dashed, upper axis); red, green and blue indicate Sh1, Sh2 and Sh3. The dashed line marks the equivalent bed height $Y^{+}=(y-H_b)^{+}=0$.
    }
    \label{fig:schematic}
\end{figure}
\par A sediment bed is prepared by randomly distributing $N_p=12\,000$ particles of diameter $D_p=H/12$ throughout the domain and allowing them to settle under gravity until the kinetic energy becomes negligible  \citep{kidanemariamInterfaceresolvedDirectNumerical2014}, where $H$ is the half-channel height. After settlement, the particles are fixed and wall turbulence is simulated over the rough bed, denoted case Sh0, as shown in figure~\ref{fig:schematic}(a). Particles satisfying criteria $F_{c,y}/[(\rho_p-\rho_f)V_pg]\geq10^{-5}$ and $|\boldsymbol{u}_p|^2/U_g^2<0.05$ are identified as bed particles \citep{schererScalingInstabilityFlat2020}, where $F_{c,y}$ is the vertical component of $\boldsymbol{F}_c$ and $U_g=\sqrt{R|\boldsymbol{g}|D_p}$ is the gravitational velocity scale. A continuous bed-height field $\eta(x,z,t)$ is obtained by Gaussian smoothing with scale $D_p$. Ensemble averaging of $\eta$ yields the equivalent bed height $H_b=\langle\overline{\eta}\rangle_{xz}+0.5D_p=3.23D_p$ and $H_f=H-H_b=8.77D_p$, where $\overline{\cdot}$ and $\langle\cdot\rangle_{xz}$ denote temporal and horizontal-plane averaging over the statistically steady state. The friction Reynolds number at $H_b$ is $\mathrm{Re}_{\tau}=236.28$. A particle-free DNS at the same $\mathrm{Re}_{\tau}$ serves as the smooth-wall baseline, denoted PF. Table~\ref{tab:pf_sh0} summarises the parameters of PF and Sh0, where $\mathrm{Re}_b=u_bH_f/\nu_f$, $u_b$ is the bulk velocity, $T_b=H_f/u_b$ is the bulk time scale, $T_{obs}$ and $T_{obs}^s$ are the total and steady-state durations, and Sh0 uses uniform grids while PF uses a non-uniform wall-normal grid.
\par Starting from Sh0, particle constraints are released and three erodible cases at different $\theta$ are realised by varying $|\boldsymbol{g}|$, denoted Sh1, Sh2 and Sh3. Their parameters, non-dimensionalized by $H_f$ and $u_{\tau}$ of Sh0, are listed in table~\ref{tab:sh1_sh2_sh3}, where the Galileo number $\mathrm{Ga}=D_p^+/\sqrt{\theta}$, the particle volume flux per unit width $q_p=\frac{\pi D_p^3}{6L_xL_z}\sum_{l}^{N_p}u_{p,x}$, and the critical Shields number $\theta_{cr}$ is computed from the empirical model of \citet{guoEmpiricalModelShields2020}. Varying only gravity ensures consistent fluid and particle properties across cases Sh0--Sh3, such as the Stokes number $\mathrm{St}^+={\rho_pD_p^2u_{\tau}^2}/({18\rho_f\nu_f})=106.02$ and $D_p^+=u_{\tau}D_p/\nu_f=26.84$. Figure~\ref{fig:schematic}(b) shows $\overline{q_p}/U_g/D_p$ versus Shields number for Sh1, Sh2 and Sh3, with dashed lines representing the fits of \citet{meyer1948formulas} and \citet{wongReanalysisCorrectionBedload2006}; the present data agree well with existing experiments. Different $\theta$ controls not only the particle transport intensity but also the dominant mode of particle motion. The inset of figure~\ref{fig:schematic}(b) shows the streamwise translational velocity $\langle \overline{u_p}\rangle_{xz}^+$ and spanwise angular velocity $\langle \overline{\omega_{p,z}}\rangle_{xz}^+$ for Sh1--Sh3 (red, green and blue). Sh1 is dominated by particles with low translational but high angular velocity, Sh3 by those with high translational but low angular velocity, and Sh2 represents an intermediate regime.

\begin{table}
    \begin{center}
    \def~{\hphantom{0}}
    \setlength{\tabcolsep}{2pt}
    \begin{tabular}{lccccccc}
    Case & $\mathrm{Re}_{\tau}$ &$\mathrm{Re}_{b}$& $[L_x\times L_y \times L_z]/H_f$ & $N_x\times N_y\times N_z$ & $\Delta t_f^+$ & $T_{obs}/T_b$ & $T_{obs}^s/T_b$\\
    PF & $ 236.28 $ &$3584.55$& $6\pi\times 1\times 2\pi$ & $512\times 128\times 512$ & $7.92\times 10^{-2}$  & $3992.67$ & $2187.76$ \\
    Sh0 & $ 236.28$ &$2606.86$& $13.63\times 1.36 \times 3.41$ & $1280\times 128\times 320$ & $5.04\times 10^{-2}$ & $752.54$ &$258.68$ \\
    \end{tabular}
    \caption{Numerical simulation parameters for the baseline cases PF and Sh0.}
    \label{tab:pf_sh0}
    \end{center}
\end{table}
\begin{table}
\begin{center}
\def~{\hphantom{0}}
\setlength{\tabcolsep}{3pt}
\begin{tabular}{lcccccccccc}
Case  & $\theta$ & $\theta_{cr}$ & $\theta/\theta_{cr}$ & $\overline{q_p}/U_g/D_p$ &  $\mathrm{Ga}$ & $\overline{\max{(\eta')}}/D_p$ & $\overline{\min{(\eta')}}/D_p$& $T_{obs}/T_b$ & $T_{obs}^s/T_b$ & \\
Sh1 & $0.0671$ & $0.0337$ & $1.99$ & $0.0361$ & $103.62$  & $0.82$ & $-0.54$ & $940.67$ & $258.68$& \\
Sh2 & $0.1840$ & $0.0339$ & $5.42$ &$0.4402$& $62.56$   & $0.71$ & $-0.83$ & $940.67$ & $258.68$& \\
Sh3 & $0.2964$ & $0.0353$ & $8.38$ &$1.4455$& $49.29$   & $0.71$ & $-0.25$ & $1175.84$& $258.68$& \\
\end{tabular}
\caption{Dimensionless characteristic parameters for the erodible cases Sh1--Sh3.}
\label{tab:sh1_sh2_sh3}
\end{center}
\end{table}

\section{Results}
\label{results}

\begin{figure}
    \centering
    \includegraphics[width=1\linewidth]{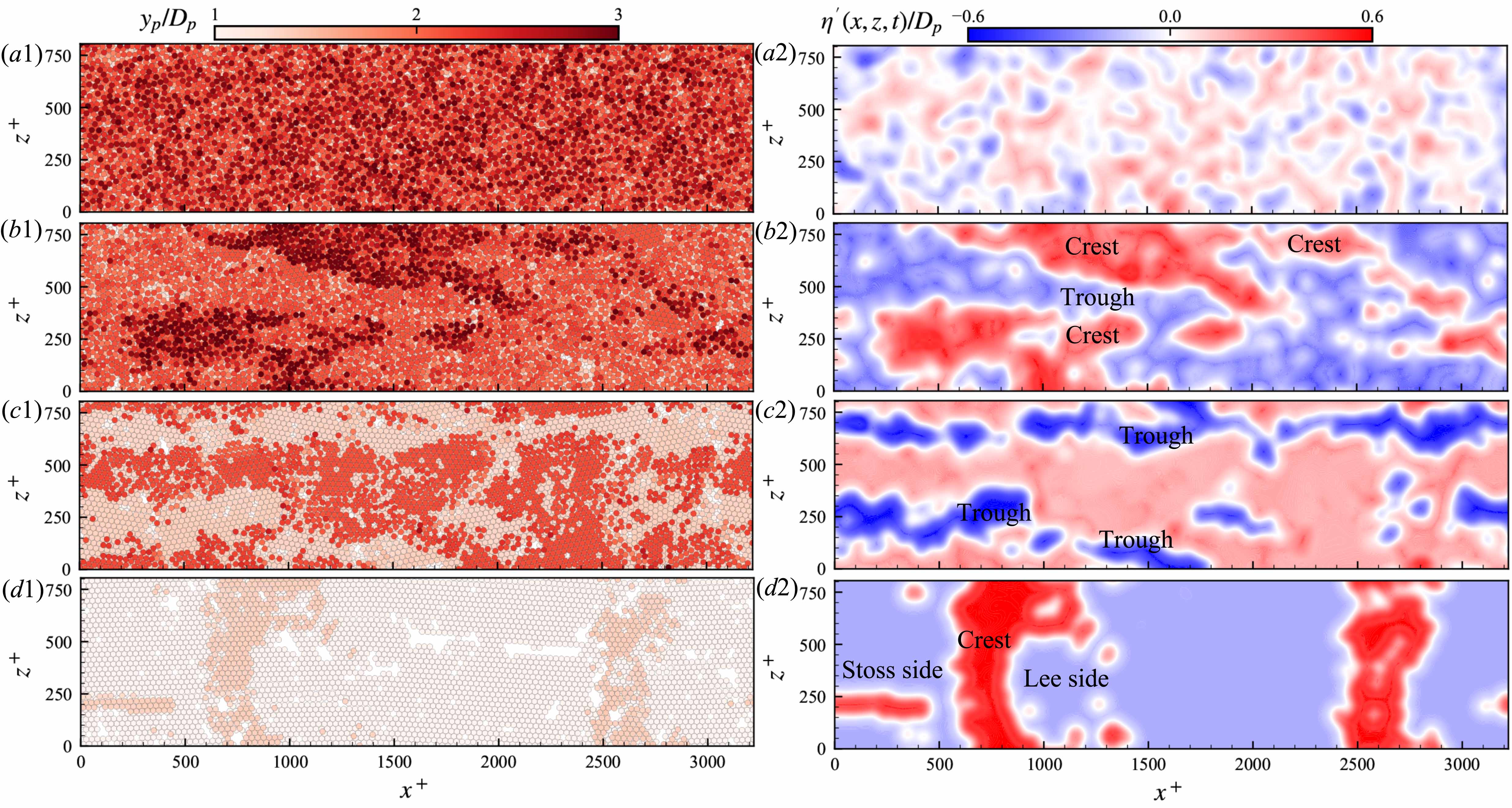}
    \caption{
    Top view of bed particles and contours of $\eta^{\prime}$ for cases Sh0-Sh3. (a1)--(d1) show instantaneous top views of particles; (a2)--(d2) show $\eta^{\prime}$ at the same time.
    }
    \label{fig:ParticleBedSurface}
\end{figure}

\noindent \par Figure~\ref{fig:ParticleBedSurface} shows instantaneous top views of the bed and contours of the bed-surface fluctuation $\eta^{\prime}=\eta(x,z,t)-\langle \eta\rangle_{xz}$. Sh0 exhibits a fragmented random $\eta^{\prime}$ pattern owing to stochastic dry settling \citep{bratbergEffectsForceDistribution2004}. Sh1 develops clear longitudinal bedforms in figure~\ref{fig:ParticleBedSurface}(b), i.e.\ streamwise sediment ridges \citep{schererRoleTurbulentLargescale2022}. Notably, \cite{kidanemariamFormationSedimentPatterns2017} found that when the streamwise domain exceeds $4H_f$ (or $75~100D_p$),  longitudinal bedforms transition to transverse bedforms. This was confirmed by \cite{zgheibDirectNumericalSimulation2018} and \citet{scherer2024turbulent} with $L_x=11.43H_f$ and $11.56H_f$, respectively: although longitudinal bedforms emerge at early times in long domains, they eventually give way to transverse bedforms at statistical steady state. Here, the domain $L_x=13.63H_f=120D_p$ exceeds the above threshold, and Sh1 has both a lower $\theta$ and longer simulation time in table~\ref{tab:sh1_sh2_sh3}, yet its longitudinal structures persist at steady state, confirming the existence of stable streamwise sediment ridges at lower $\theta$. Sh2 retains longitudinal bedforms in figure~\ref{fig:ParticleBedSurface}(c). However, whereas the longitudinal bedforms in Sh1 are dominated by red crests (positive $\eta'$), those in Sh2 are dominated by blue troughs (negative $\eta'$); the maximum crest $\overline{\max{(\eta')}}/D_p$ and maximum trough $\overline{\min{(\eta')}}/D_p$ are listed in table~\ref{tab:sh1_sh2_sh3}. This indicates that the former primarily results from particle deposition, while the latter is primarily driven by erosion. We refer to these as streamwise erosion trenches in Sh2. Sh3 transitions to transverse ripples in figure~\ref{fig:ParticleBedSurface}(d), consistent with \citet{schererScalingInstabilityFlat2020}. These results reveal for the first time that within $\theta\in[0.0671,0.2964]$, bedforms undergo a complete longitudinal-to-transverse transition: from sediment ridges, through erosion trenches, to ripples.

\begin{figure}
    \centering
    \includegraphics[width=1\linewidth]{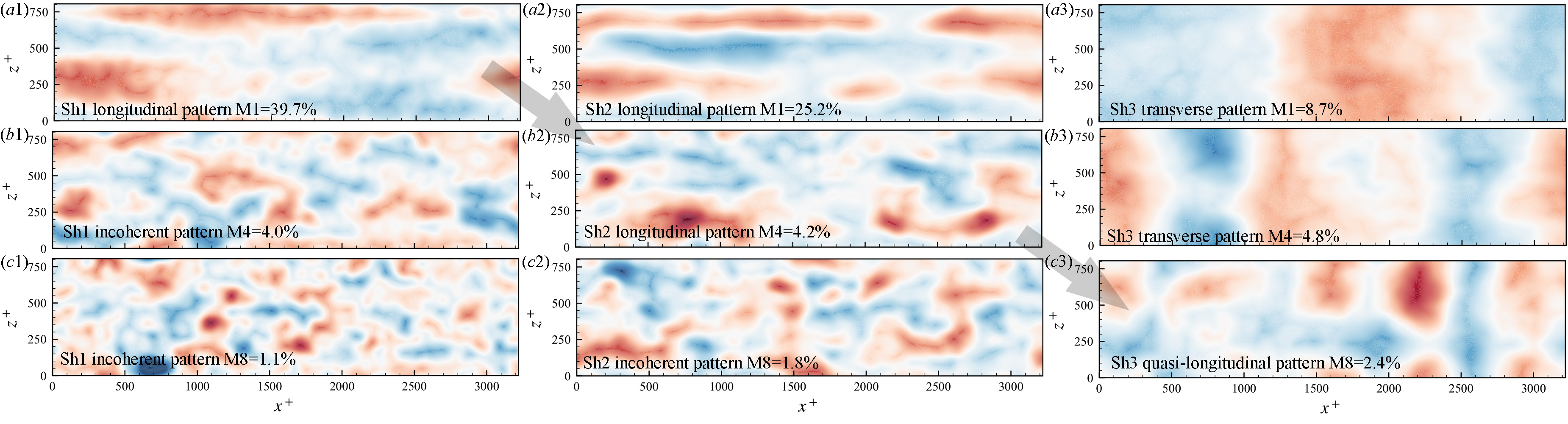}
    \caption{
    POD of $\eta^{\prime}$ for Sh1-Sh3. (a1)--(c1) show modes 1, 4 and 8 for Sh1 with percentage contributions. (a2)--(c2) and (a3)--(c3) show the corresponding results for Sh2 and Sh3. Grey arrows indicate the ``top-left to bottom-right'' transfer of modal weight. The color scale as in figure\ref{fig:ParticleBedSurface}(a2-d2).
    }
    \label{fig:Mod_Surface_All_text}
\end{figure}

\par We performed proper orthogonal decomposition (POD) on $\eta^{\prime}$, and figure~\ref{fig:Mod_Surface_All_text} displays the most representative modes (1, 4 and 8). In Sh1, mode 1 ($\text{M1}=39.7\%$) corresponds to streamwise sediment ridges, which is the longitudinal pattern, while modes 4 and 8 ($\text{M4}=4.0\%$, $\text{M8}=1.1\%$) exhibit incoherent patterns. At this low $\theta$, fluid shear is insufficient to break intra-bed force chains and the streamwise ridges remain stable. In Sh2, the modal contributions are $\text{M1}=25.2\%$, $\text{M4}=4.2\%$ and $\text{M8}=1.8\%$; the bedform remains longitudinal but the leading mode weakens. This indicates that as $\theta$ increases, erosion intensifies, local force chains are disrupted, and the modal energy disperses across multiple higher-order longitudinal modes. In Sh3, $\text{M1}=8.7\%$, $\text{M4}=4.8\%$, $\text{M8}=2.4\%$. Here, modes 1 and 4 have transitioned to transverse bedform, yet mode 8 retains a fragmented quasi-longitudinal form. Along the diagonal from figure~\ref{fig:Mod_Surface_All_text}(a1) to figure~\ref{fig:Mod_Surface_All_text}(c3), a clear transfer pathway emerges: longitudinal structures at low $\theta$, though supplanted by transverse bedforms at high $\theta$, persist as remnants in higher-order modes.
We term this the 'top-left to bottom-right' pathway, revealing that large-scale bedforms in low-order modes are preferentially reshaped by saltation, whereas small-scale longitudinal features survive in higher-order modes as residual imprints of the transition.

\begin{figure}
  \centerline{\includegraphics[width=1\linewidth]{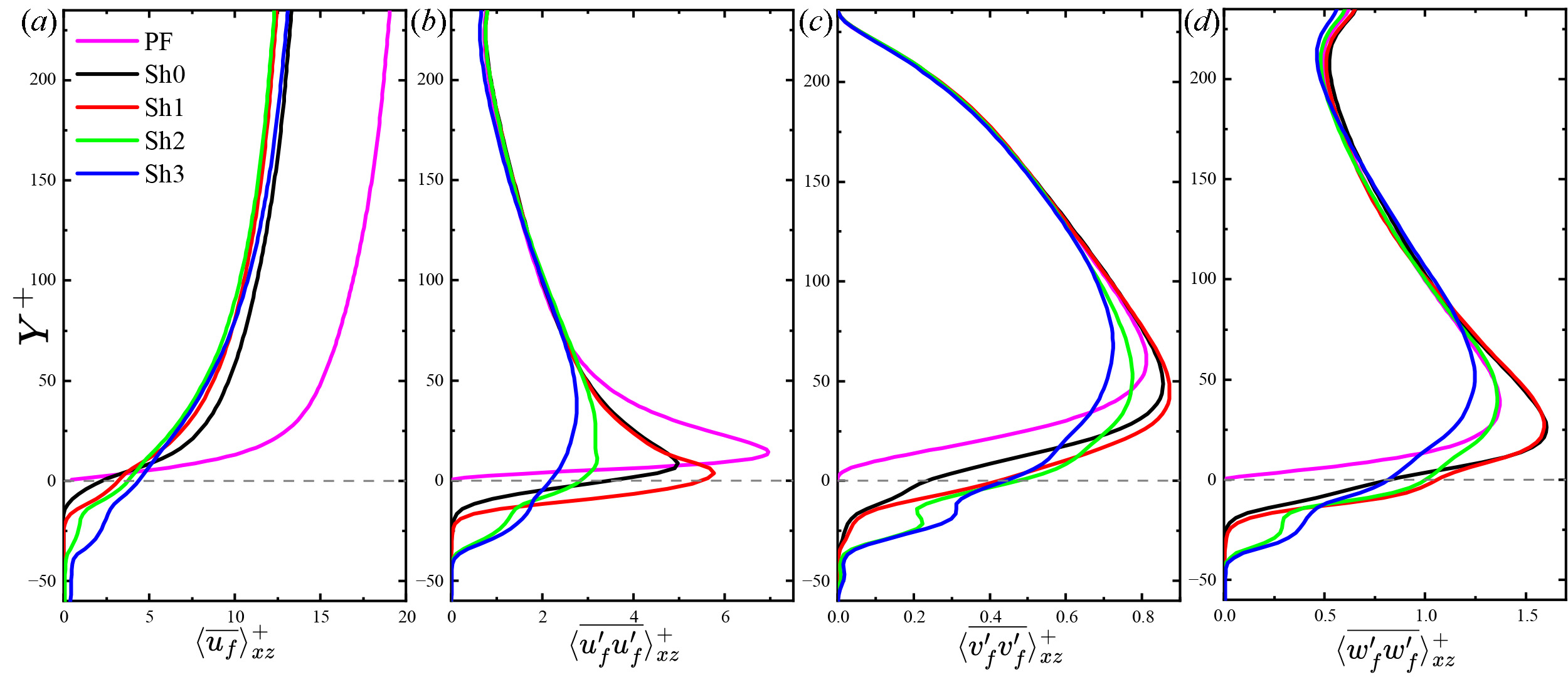}}%
  \caption{Wall-normal profiles of fluid velocity.
    (a) Mean streamwise $\langle \overline{u_f}\rangle_{xz}^+$.
    (b) Streamwise fluctuation $\langle \overline{u_f^{\prime}u_f^{\prime}}\rangle_{xz}^+$.
    (c) Wall-normal fluctuation $\langle \overline{v_f^{\prime}v_f^{\prime}}\rangle_{xz}^+$.
    (d) Spanwise fluctuation $\langle \overline{w_f^{\prime}w_f^{\prime}}\rangle_{xz}^+$.
    }
  \label{fig:profile}
\end{figure}

\par The Shields number significantly alters bed-particle motion, affecting the mean velocity and turbulence statistics.
Figure~\ref{fig:profile}(a) shows the mean streamwise fluid velocity $\langle \overline{u_f}\rangle_{xz}^+$.
In Sh0, $\langle \overline{u_f}\rangle_{xz}^+$ is substantially reduced relative to PF above the bed, while the non-zero $\langle \overline{u_f}\rangle_{xz}$ within the bed indicates weak percolation.
From Sh1 to Sh3, particle motion enhances percolation within the bed, so $\langle \overline{u_f}\rangle_{xz}^+$ ($Y^+<0$) increases monotonically with $\theta$.
Above the bed, particle entrainment enhances fluid--particle momentum exchange, causing the fluid velocity in Sh1 and Sh2 to decrease with $\theta$.
Although Sh3 shows further attenuation of $\langle \overline{u_f}\rangle_{xz}^+$ in the near-wall region, it surpasses Sh1 and Sh2 for $Y^+>100$, resembling the non-monotonic variation of bedforms with flow intensity reported in field observations \citep{van2007unified}: bedform grow at low transport intensity but is washed out at high transport intensity.
Consistently, the peak-to-trough amplitude of $\eta'$ in table~\ref{tab:sh1_sh2_sh3} shows that Sh2 has the largest amplitude, whereas Sh3 has the smallest.

\begin{figure}
    \centering
    \includegraphics[width=1\linewidth]{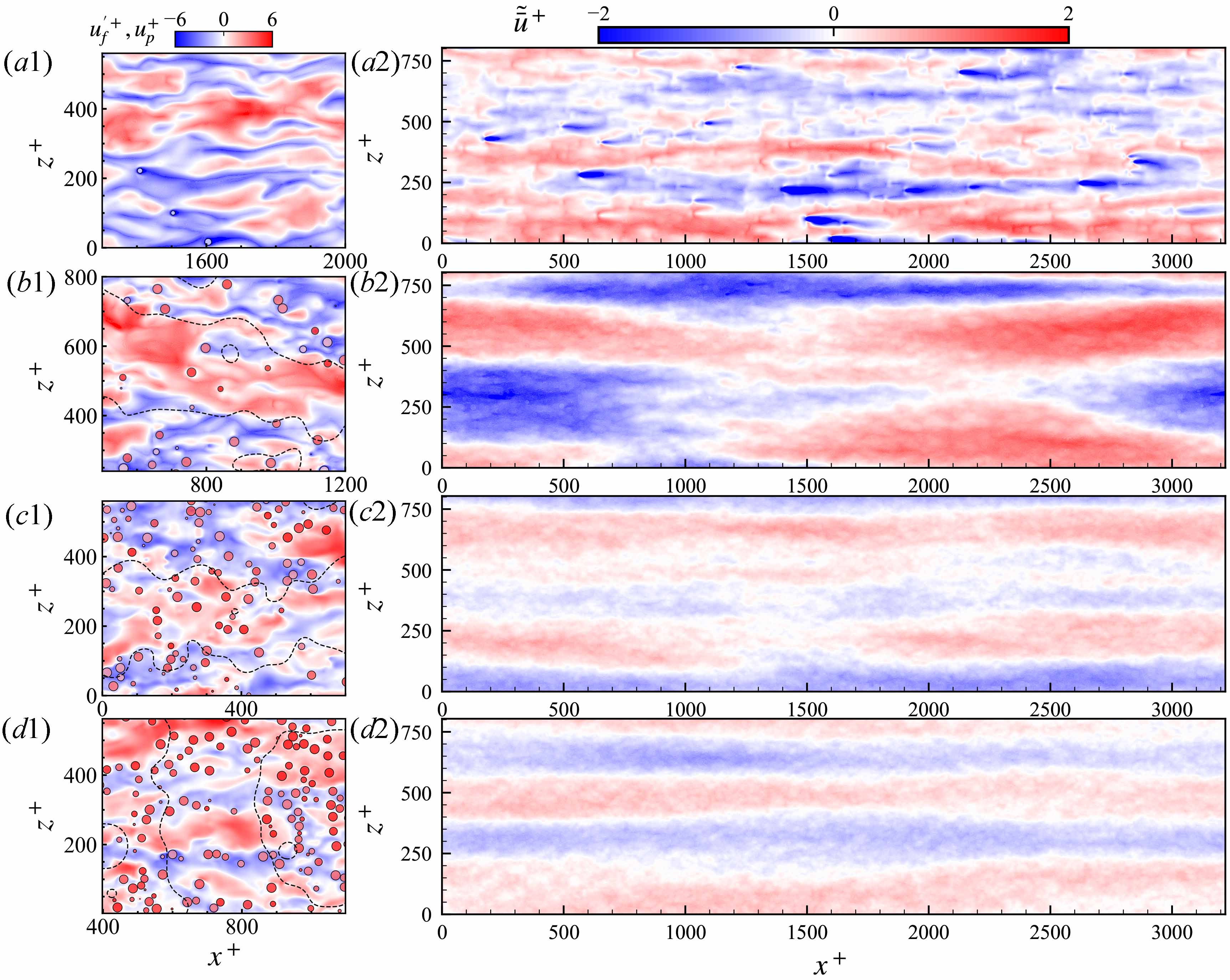}
    \caption{
    (a1)--(d1): Instantaneous $u_f^{\prime+}$ at $Y^+=16$, corresponding to the same instant as figure~\ref{fig:ParticleBedSurface}.
    Black dashed lines mark the $\eta'/D_p=0$ contour from figure~\ref{fig:ParticleBedSurface}.
    (a2)--(d2): Contours of $\widetilde{\bar{u}}$ at $Y^+=16$ over $T_{obs}^{s}$.
    }
    \label{fig:snapshot_uf_xz}
\end{figure}

\par Figures~\ref{fig:profile}(b--d) present the velocity fluctuation profiles. Relative to PF, the fixed-rough case Sh0 exhibits a suppressed peak in streamwise fluctuations $\langle \overline{u_f^{\prime}u_f^{\prime}}\rangle_{xz}^+$, while the peaks of wall-normal $\langle \overline{v_f^{\prime}v_f^{\prime}}\rangle_{xz}^+$ and spanwise $\langle \overline{w_f^{\prime}w_f^{\prime}}\rangle_{xz}^+$ fluctuations are enhanced. The rough bed disrupts the near-wall quasi-streamwise vortices and high- and low-speed streaks of the smooth-wall flow, redistributing TKE among the three components \citep{jimenezTurbulentFlowsRough2004}.
In Sh1, the peak of $\langle \overline{u_f^{\prime}u_f^{\prime}}\rangle_{xz}^+$ lies below that of PF, consistent with \citet{schererRoleTurbulentLargescale2022}; however, the peaks of $\langle \overline{u_f^{\prime}u_f^{\prime}}\rangle_{xz}^+$ and $\langle \overline{v_f^{\prime}v_f^{\prime}}\rangle_{xz}^+$ both exceed those of Sh0, while the peak of $\langle \overline{w_f^{\prime}w_f^{\prime}}\rangle_{xz}^+$ is comparable. This contrasts with \citet{jiDirectNumericalSimulation2013}, who reported that the peak fluctuations in all components were instead attenuated at similar $\theta/\theta_{cr}$. The discrepancy may arise from the smaller particle number and shorter streamwise domain in \citet{jiDirectNumericalSimulation2013}, which precluded the formation of pronounced bedforms.
Notably, \citet{vowinckelFluidParticleInteraction2014} observed significant enhancement of fluid fluctuations across the entire boundary layer for both ridge-like and dune-like bedforms, whereas in the present cases negligible differences appear above $Y^+>100$. This discrepancy likely stems from the fact that in \citet{vowinckelFluidParticleInteraction2014} the bottom bed layer was artificially fixed and only the particles above it could move freely---i.e., the bed was not truly erodible.
In Sh2 and Sh3, the fluctuation peaks in all three directions are suppressed relative to both PF and Sh0, and this suppression is more pronounced than in \citet{jiDirectNumericalSimulation2013}, primarily owing to the higher Shields numbers of Sh2 and Sh3.
The dune-like (transverse) bedforms in \citet{vowinckelFluidParticleInteraction2014} were achieved only by artificially reducing the number of mobile particles on the fixed bed. By contrast, in the erodible Sh3, which also exhibits transverse structures, the stronger transport leads to more mobile particles above the bed, and consequently the near-wall fluctuations are suppressed relative to Sh0.

\par To probe the physical mechanism underlying the non-monotonic turbulence modulation, figure~\ref{fig:snapshot_uf_xz}(a1)--(d1) shows enlarged views of instantaneous $u_f^{\prime+}$ snapshots in the $x$--$z$ plane at $Y^+=16$, corresponding to the same time as figure~\ref{fig:ParticleBedSurface}. The time-averaged velocity $\overline{u_f}$ over the entire statistically steady period $T_{obs}^{s}$ is used to compute the spatial fluctuation field $\widetilde{\bar{u}}=\overline{u_f}-\langle \overline{u_f}\rangle_{xz}$, i.e.\ the form-induced velocity, shown in figures~\ref{fig:snapshot_uf_xz}(a2)--(d2).
Figure~\ref{fig:snapshot_uf_xz}(a1) shows that in Sh0, wakes induced by individual fixed particles are embedded within the near-wall high- and low-speed streaks; these wake structures give rise to pronounced negative form-induced velocity in figure~\ref{fig:snapshot_uf_xz}(a2).
Although high- and low-speed streaks are also clearly present in figure~\ref{fig:snapshot_uf_xz}(b1), they differ markedly from Sh0.
The most direct evidence is the strong spatial correspondence between the fluid velocity distribution and the bed morphology.
In figure~\ref{fig:snapshot_uf_xz}(b1), red high-speed streaks coincide with troughs of the longitudinal bed, while blue low-speed regions coincide with crests; we refer to these as topography-conditioned streaks.
These topography-conditioned streaks persist in Sh2 (figure~\ref{fig:snapshot_uf_xz}c1). However, owing to stronger particle transport, the velocity field is simultaneously contaminated by single-particle wakes, particle-cluster wakes and topography-conditioned streaks, causing the latter to be significantly weakened (figure~\ref{fig:snapshot_uf_xz}c2).
In figure~\ref{fig:snapshot_uf_xz}(d1), the correlation between the $\eta'=0$ contour (black dashed) and the velocity field has largely vanished. In the range $600<x^+<900$, particles are sparsely distributed over the ripple crests, whereas pronounced particle accumulation occurs in the lee-side region for $x^+>900$.
In figure~\ref{fig:snapshot_uf_xz}(d2), the form-induced velocity is further reduced. This non-monotonic modulation originates from the competition between static bed particles and near-wall moving particles. At low $\theta$, particles accumulated on the bed surface induce form-induced velocity streaks near the wall, enhancing turbulent fluctuations; at high $\theta$, the increase of saltating particles in the surface layer instead suppresses those fluctuations.
\begin{figure}
    \centering
    \includegraphics[width=1\linewidth]{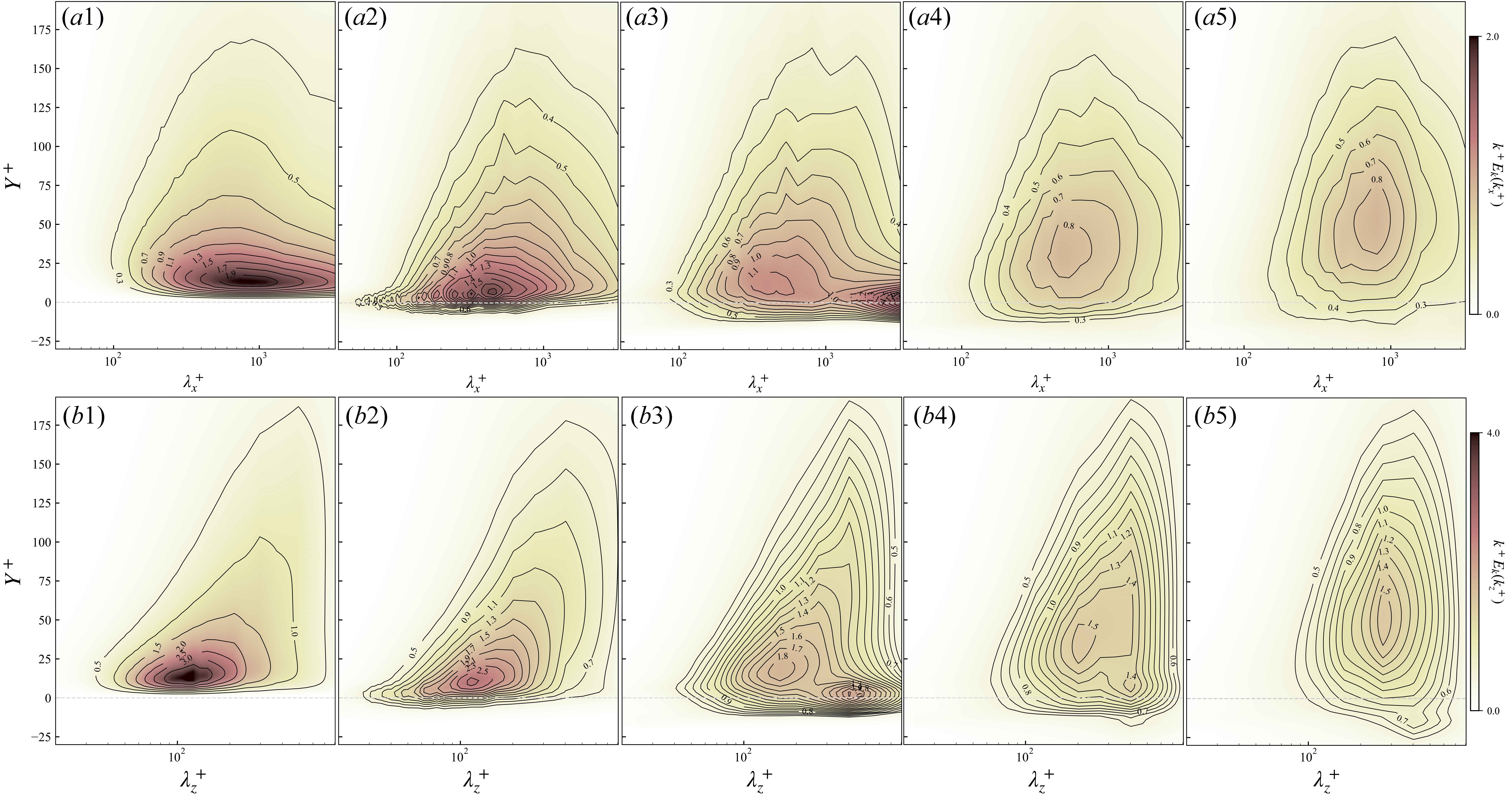}
    \caption{
    One-dimensional premultiplied energy spectra versus $Y^+$. Left column: streamwise; right column: spanwise. Panels (a1,b1): PF; (a2,b2): Sh0; (a3,b3): Sh1; (a4,b4): Sh2; (a5,b5): Sh3.
    }
    \label{fig:premultiplied_spectra}
\end{figure}

\par Figures~\ref{fig:premultiplied_spectra}(a1,b1) show the PF spectra; an artificial blank region for $Y^+<0$ maintains a consistent abscissa scale with Sh0. For Sh0 in figures~\ref{fig:premultiplied_spectra}(a2,b2), the streamwise peak is $(k_x^+ E_{uu}^+)_{\max}=1.80$ at $Y^+=6.07$ ($\lambda_x^+=460$), and the spanwise peak is $2.77$ at $Y^+=11.10$ ($\lambda_z^+=115$). The fixed particle bed thus significantly suppresses the scale and intensity of near-wall streamwise fluctuations \citep{flackRoughnessEffectsWallbounded2014}, surface roughness enhances small-scale fluctuations while disrupting large-scale streak continuity, systematically shortening the streamwise wavelength. Very weak local secondary peaks near the particle wall are attributed to multi-scale wakes induced by individual particles (see figure~\ref{fig:snapshot_uf_xz}a). Figures~\ref{fig:premultiplied_spectra}(a3,b3) reveal that the streamwise depositional ridges generate a distinct near-wall secondary spectral peak---the \textit{form-induced second peak}. In Sh1, the conventional first peak is $\max(k^+E^+)=1.19$ at $Y^+=13$ ($\lambda_x^+=395$) in the streamwise direction and $1.89$ at $Y^+=18$ ($\lambda_z^+=161$) in the spanwise direction. Remarkably, the form-induced second peak reaches $1.85$ at $Y^+=1$ ($\lambda_x^+=3220$) streamwise and $2.02$ at $Y^+=3$ ($\lambda_z^+=377$) spanwise, exceeding the conventional turbulent first peak in both directions. With increasing $\theta$, the streamwise form-induced second peak vanishes completely in Sh2 and Sh3. In the spanwise direction, enhanced saltation and increased saltation-layer thickness cause the two peaks to separate: the second peak descends to $Y^+=10$ and $Y^+=-12$ for Sh2 and Sh3, decaying to $1.38$ and $0.75$. Notably, the streamwise scale of the form-induced velocity structure expands with increasing $\theta$, nearly spanning the entire domain in Sh3. At low $\theta$, particle velocities are relatively low ($Y^+<0$ in figure~\ref{fig:schematic}b), and slow bed evolution preserves the temporal phase coherence of the form-induced velocity in the time-averaged field; as $\theta$ increases, intensified transport eliminates the streamwise phase difference through time averaging, leaving only spanwise anisotropy. Thus, the emergence and subsequent decay of this form-induced second peak constitute the core mechanism of the non-monotonic TKE modulation.

\section{Conclusion}
\noindent Using PR-DNS, we investigate the non-monotonic modulation of near-wall turbulence over an erodible bed relative to a fixed rough bed by varying θ. This modulation is tightly coupled to the bedform transition: as $\theta$ increases, the bed evolves from streamwise sediment ridges through erosion trenches to transverse ripples, with stable longitudinal bedforms persisting at lower $\theta$ than previously reported. POD reveals a `top-left to bottom-right' redistribution of modal contribution from a single dominant mode to higher-order modes; quasi-longitudinal remnants in these higher-order modes identify saltating particles as the primary agents disrupting coherent longitudinal bedforms. At low $\theta$, the longitudinal ridges generate form-induced streaks and a form-induced second peak at $Y^{+}\approx5$ that exceeds the conventional near-wall turbulent peak, enhancing TKE relative to Sh0. As $\theta$ increases, intensified saltation disrupts these coherent structures: the streamwise secondary peak vanishes, the spanwise secondary peak weakens below the near-wall first peak, and TKE is suppressed. These findings directly link bedform evolution to turbulence modulation in the erodible system.

\backsection[Funding]{This work is supported by the National Natural Science Foundation of China (No. 12388101), the Natural Science Foundation of Gansu Province (No. 25JRRA636) and the National Numerical Wind Tunnel project.
The authors thank the Fundamental and Interdisciplinary Disciplines Breakthrough Plan of the Ministry of Education of China (No. JYB2025XDXM910) for supporting this work, and the Supercomputing Center of Lanzhou University for providing computational resources.}

\backsection[Declaration of interests]{The authors report no conflict of interest.}
\bibliographystyle{jfm}
\bibliography{references}

\begin{thebibliography}{30}
\expandafter\ifx\csname natexlab\endcsname\relax\def\natexlab#1{#1}\fi
\def\au#1{#1} \def\ed#1{#1} \def\yr#1{#1}\def\at#1{#1}\def\jt#1{\textit{#1}}
  \def\bt#1{#1}\def\bvol#1{\textbf{#1}} \def\vol#1{#1} \def\pg#1{#1}
  \def\publ#1{#1}\def\arxiv#1{#1}\def\org#1{#1}\def\st#1{\textit{#1}}

\bibitem[Bennett {\em et~al.\/}(1998)Bennett, Bridge \&
  Best]{bennettFluidSedimentDynamics1998}
{\sc \au{Bennett, Sean~J.}, \au{Bridge, John~S.} \& \au{Best, James~L.}}
  \yr{1998}  \at{Fluid and sediment dynamics of upper stage plane beds}.
  \jt{Journal of Geophysical Research: Oceans}  \bvol{103}~(C1),
  \pg{1239--1274}, \_eprint:
  https://agupubs.onlinelibrary.wiley.com/doi/pdf/10.1029/97JC02764.

\bibitem[Biegert {\em et~al.\/}(2017)Biegert, Vowinckel \&
  Meiburg]{biegertCollisionModelGrainresolving2017}
{\sc \au{Biegert, Edward}, \au{Vowinckel, Bernhard} \& \au{Meiburg, Eckart}}
  \yr{2017}  \at{A collision model for grain-resolving simulations of flows
  over dense, mobile, polydisperse granular sediment beds}.  \jt{Journal of
  Computational Physics}  \bvol{340},  \pg{105--127}.

\bibitem[Brandt \& Coletti(2022)]{brandtParticleLadenTurbulenceProgress2022}
{\sc \au{Brandt, Luca} \& \au{Coletti, Filippo}} \yr{2022}
  \at{Particle-{Laden} {Turbulence}:{Progress} and {Perspectives}}.  \jt{Annual
  Review of Fluid Mechanics}  \bvol{54}~(1),  \pg{159--189}, \_eprint:
  https://doi.org/10.1146/annurev-fluid-030121-021103.

\bibitem[Bratberg(2004)]{bratbergEffectsForceDistribution2004}
{\sc \au{Bratberg, Ivar}} \yr{2004}  \bt{Effects of the force distribution in
  dry granular materials}.

\bibitem[Breugem(2012)]{breugemSecondorderAccurateImmersed2012}
{\sc \au{Breugem, Wim-Paul}} \yr{2012}  \at{A second-order accurate immersed
  boundary method for fully resolved simulations of particle-laden flows}.
  \jt{Journal of Computational Physics}  \bvol{231}~(13),  \pg{4469--4498}.

\bibitem[Dey(2014)]{dey2014fluvial}
{\sc \au{Dey, Subhasish}} \yr{2014} {\em Fluvial hydrodynamics\/}.
  \publ{Springer}.

\bibitem[Flack \& Schultz(2014)]{flackRoughnessEffectsWallbounded2014}
{\sc \au{Flack, Karen~A.} \& \au{Schultz, Michael~P.}} \yr{2014}  \at{Roughness
  effects on wall-bounded turbulent flows}.  \jt{Physics of Fluids}
  \bvol{26}~(10),  \pg{101305}.

\bibitem[Garcia(2008)]{garcia2008sedimentation}
{\sc \au{Garcia, Marcelo}} \yr{2008}  \bt{Sedimentation engineering: processes,
  measurements, modeling, and practice}.  \publ{American Society of Civil
  Engineers}.

\bibitem[Guo(2020)]{guoEmpiricalModelShields2020}
{\sc \au{Guo, Junke}} \yr{2020}  \at{Empirical model for shields diagram and
  its applications}.  \jt{Journal of Hydraulic Engineering}  \bvol{146}~(6),
  \pg{4020038}.

\bibitem[Ji {\em et~al.\/}(2013)Ji, Munjiza, Avital, Ma \&
  Williams]{jiDirectNumericalSimulation2013}
{\sc \au{Ji, C.}, \au{Munjiza, A.}, \au{Avital, E.}, \au{Ma, J.} \&
  \au{Williams, J. J.~R.}} \yr{2013}  \at{Direct numerical simulation of
  sediment entrainment in turbulent channel flow}.  \jt{Physics of Fluids}
  \bvol{25}~(5),  \pg{056601}.

\bibitem[Jiménez(2004)]{jimenezTurbulentFlowsRough2004}
{\sc \au{Jiménez, Javier}} \yr{2004}  \at{Turbulent flows over rough walls}.
  \jt{Annual Review of Fluid Mechanics, Vol 43}  \bvol{36}~(1),  \pg{173--196}.

\bibitem[Kidanemariam \&
  Uhlmann(2014)]{kidanemariamInterfaceresolvedDirectNumerical2014}
{\sc \au{Kidanemariam, Aman~G.} \& \au{Uhlmann, Markus}} \yr{2014}
  \at{Interface-resolved direct numerical simulation of the erosion of a
  sediment bed sheared by laminar channel flow}.  \jt{International Journal of
  Multiphase Flow}  \bvol{67},  \pg{174--188}.

\bibitem[Kidanemariam \&
  Uhlmann(2017)]{kidanemariamFormationSedimentPatterns2017}
{\sc \au{Kidanemariam, Aman~G.} \& \au{Uhlmann, Markus}} \yr{2017}
  \at{Formation of sediment patterns in channel flow: minimal unstable systems
  and their temporal evolution}.  \jt{Journal of Fluid Mechanics}  \bvol{818},
  \pg{716--743}.

\bibitem[Li \&
  McKenna~Neuman(2012)]{liBoundarylayerTurbulenceCharacteristics2012a}
{\sc \au{Li, Bailiang} \& \au{McKenna~Neuman, Cheryl}} \yr{2012}
  \at{Boundary‐layer turbulence characteristics during aeolian saltation}.
  \jt{Geophysical Research Letters}  \bvol{39}~(11),  \pg{2012GL052234}.

\bibitem[Liu {\em et~al.\/}(2022)Liu, Feng \&
  Zheng]{liuExperimentalInvestigationEffects2022}
{\sc \au{Liu, Hongyou}, \au{Feng, Yuen} \& \au{Zheng, Xiaojing}} \yr{2022}
  \at{Experimental investigation of the effects of particle near-wall motions
  on turbulence statistics in particle-laden flows}.  \jt{Journal of Fluid
  Mechanics}  \bvol{943},  \pg{A8}.

\bibitem[Meyer-Peter \& Müller(1948)]{meyer1948formulas}
{\sc \au{Meyer-Peter, E.} \& \au{Müller, R.}} \yr{1948} Formulas for bed-load
  transport.  \bt{In {\em Proceedings of the 2nd {Meeting} of the
  {International} {Association} for {Hydraulic} {Structures} {Research}\/}},
  \pg{pp. 39--64}.  \publ{Stockholm, Sweden: International Association for
  Hydraulic Research}.

\bibitem[Raudkivi(1963)]{raudkiviStudySedimentRipple1963}
{\sc \au{Raudkivi, A.~J.}} \yr{1963}  \at{Study of sediment ripple formation}.
  \jt{Journal of the Hydraulics Division}  \bvol{89}~(6),  \pg{15--34}.

\bibitem[Revil-Baudard {\em et~al.\/}(2016)Revil-Baudard, Chauchat, Hurther \&
  Eiff]{revil-baudardTurbulenceModificationsInduced2016}
{\sc \au{Revil-Baudard, T.}, \au{Chauchat, J.}, \au{Hurther, D.} \& \au{Eiff,
  O.}} \yr{2016}  \at{Turbulence modifications induced by the bed mobility in
  intense sediment-laden flows}.  \jt{Journal of Fluid Mechanics}  \bvol{808},
  \pg{469--484}.

\bibitem[Scherer {\em et~al.\/}(2020)Scherer, Kidanemariam \&
  Uhlmann]{schererScalingInstabilityFlat2020}
{\sc \au{Scherer, Markus}, \au{Kidanemariam, Aman~G.} \& \au{Uhlmann, Markus}}
  \yr{2020}  \at{On the scaling of the instability of a flat sediment bed with
  respect to ripple-like patterns}.  \jt{Journal of Fluid Mechanics}
  \bvol{900},  \pg{A1}.

\bibitem[Scherer {\em et~al.\/}(2022)Scherer, Uhlmann, Kidanemariam \&
  Krayer]{schererRoleTurbulentLargescale2022}
{\sc \au{Scherer, Markus}, \au{Uhlmann, Markus}, \au{Kidanemariam, Aman~G.} \&
  \au{Krayer, Michael}} \yr{2022}  \at{On the role of turbulent large-scale
  streaks in generating sediment ridgesa}.  \jt{Journal of Fluid Mechanics}
  \bvol{930},  \pg{A11}.

\bibitem[Scherer(2024)]{scherer2024turbulent}
{\sc \au{Scherer, Markus~Maximilian}} \yr{2024} {\em Turbulent coherent
  structures, secondary currents and sediment ridges\/}.  \publ{KIT Scientific
  Publishing}.

\bibitem[Shields {\em et~al.\/}(1936)Shields, Ott \&
  Van~Uchelen]{shields1936application}
{\sc \au{Shields, Albert}, \au{Ott, Wp} \& \au{Van~Uchelen, Jc}} \yr{1936}
  \at{Application of similarity principles and turbulence research to bed-load
  movement} .

\bibitem[Van~Rijn(2007)]{van2007unified}
{\sc \au{Van~Rijn, Leo~C}} \yr{2007}  \at{Unified view of sediment transport by
  currents and waves. {III}: graded beds}.  \jt{Journal of Hydraulic
  Engineering}  \bvol{133}~(7),  \pg{761--775}.

\bibitem[Vowinckel {\em et~al.\/}(2014)Vowinckel, Kempe \&
  Fröhlich]{vowinckelFluidParticleInteraction2014}
{\sc \au{Vowinckel, Bernhard}, \au{Kempe, Tobias} \& \au{Fröhlich, Jochen}}
  \yr{2014}  \at{Fluid–particle interaction in turbulent open channel flow
  with fully-resolved mobile beds}.  \jt{Advances in Water Resources}
  \bvol{72},  \pg{32--44}.

\bibitem[Wong \& Parker(2006)]{wongReanalysisCorrectionBedload2006}
{\sc \au{Wong, Miguel} \& \au{Parker, Gary}} \yr{2006}  \at{Reanalysis and
  correction of bed-load relation of meyer-peter and müller using their own
  database}.  \jt{Journal of Hydraulic Engineering}  \bvol{132}~(11),
  \pg{1159--1168}.

\bibitem[Yalin(1977)]{yalinMechanicsSedimentTransport1977a}
{\sc \au{Yalin, M.~S.}} \yr{1977}  \at{Mechanics of {Sediment} {Transport}} .

\bibitem[Zgheib {\em et~al.\/}(2018)Zgheib, Fedele, Hoyal, Perillo \&
  Balachandar]{zgheibDirectNumericalSimulation2018}
{\sc \au{Zgheib, N.}, \au{Fedele, J.~J.}, \au{Hoyal, D. C. J.~D.}, \au{Perillo,
  M.~M.} \& \au{Balachandar, S.}} \yr{2018}  \at{Direct numerical simulation of
  transverse ripples: 1. {Pattern} initiation and bedform interactions}.
  \jt{Journal of Geophysical Research: Earth Surface}  \bvol{123}~(3),
  \pg{448--477}.

\bibitem[Zhang {\em et~al.\/}(2008)Zhang, Wang \&
  Lee]{zhangSimultaneousPIVPTV2008}
{\sc \au{Zhang, Wei}, \au{Wang, Yuan} \& \au{Lee, Sang~Joon}} \yr{2008}
  \at{Simultaneous {PIV} and {PTV} measurements of wind and sand particle
  velocities}.  \jt{Experiments in Fluids}  \bvol{45}~(2),  \pg{241--256}.

\bibitem[Zheng {\em et~al.\/}(2021)Zheng, Feng \&
  Wang]{zhengModulationTurbulenceSaltating2021}
{\sc \au{Zheng, Xiaojing}, \au{Feng, Shengjun} \& \au{Wang, Ping}} \yr{2021}
  \at{Modulation of turbulence by saltating particles on erodible bed surface}.
   \jt{Journal of Fluid Mechanics}  \bvol{918},  \pg{A16}.

\bibitem[Zhu {\em et~al.\/}(2022)Zhu, Hu, Lei, Shen \&
  Zheng]{zhuParticleResolvedSimulation2022}
{\sc \au{Zhu, Zhengping}, \au{Hu, Ruifeng}, \au{Lei, Yinghaonan}, \au{Shen,
  Lian} \& \au{Zheng, Xiaojing}} \yr{2022}  \at{Particle resolved simulation of
  sediment transport by a hybrid parallel approach}.  \jt{International Journal
  of Multiphase Flow}  \bvol{152},  \pg{104072}.

\end{thebibliography}
\end{document}